\documentclass[prl,10pt,twocolumn,superscriptaddress,showpacs]{revtex4}
\usepackage{amsmath}
\usepackage{latexsym}
\usepackage{amssymb}
\usepackage{graphics,epstopdf}
\usepackage{graphicx}
\usepackage[colorlinks=true, citecolor=blue, urlcolor=blue ]{hyperref}
\usepackage{float}
\usepackage{graphicx}
\usepackage{amsfonts}

\begin{document}

\title{Necessary and sufficient state condition for two-qubit steering using two measurement settings
per party and monogamy of steering}

\author{Shiladitya Mal}
\email{shiladitya.27@gmail.com}
\affiliation{S. N. Bose National Centre for Basic Sciences, Block JD, Sector III, Salt Lake, Kolkata 700 098, India}

\author{Debarshi Das}
\email{debarshidas@jcbose.ac.in}
\affiliation{Center for Astroparticle Physics and Space Science (CAPSS),
Bose Institute, Block EN, Sector V, Salt Lake, Kolkata 700 091, India}

\author{Souradeep Sasmal}
\email{souradeep@mail.jcbose.ac.in}
\affiliation{Center for Astroparticle Physics and Space Science (CAPSS),
Bose Institute, Block EN, Sector V, Salt Lake, Kolkata 700 091, India}

\author{A. S. Majumdar}
\email{archan@bose.res.in}
\affiliation{S. N. Bose National Centre for Basic Sciences, Block JD, Sector III, Salt Lake, Kolkata 700 098, India}

\begin{abstract}
We consider the Cavalcanti-Foster-Fuwa-Wiseman inequality~\cite{achsh} which is a necessary and
sufficient steerability condition for two-qubit states with two measurement settings on each side.
We derive the criterion which an arbitrary two-qubit state must satisfy in order to violate
this inequality, and obtain its maximum attainable violation in quantum mechanics. The derived
condition on the state parameters enables us to establish a tight monogamy relation for
two-qubit steering.  
\end{abstract} 


\maketitle

\section{I. INTRODUCTION} 

Quantum correlations are incompatible with classical mechanics. In the seminal paper, Einstein-Podolsky-Rosen (EPR)~\cite{EPR} presented an argument sowing the seeds of incompatibility between a classical-like local realist description and quantum theory. Later,  Bell formulated a quantitative
bound for local realist correlations, that was shown to be violated by quantum mechanics~ \cite{bell}.
In response to the EPR paper,  Schr\"odinger showed that quantum correlations enabled an observer to steer a system which is not in her possession~\cite{sr}. Subsequently, Reid proposed a criterion for demonstrating  the EPR paradox~\cite{reid} using the Heisenberg uncertainty relation. 

The concept of steering in the form of an information theoretic task was introduced by Wiseman
{\it et al.}~\cite{wiseman, wiseman2}. They showed that steering occurs due to the lack of a local hidden state (LHS) model, making it impossible for an observer to simulate classically the state of another remote party. Cavalcanti {\it et al}.~\cite{cvl} have established the link between the demonstration of the EPR argument as formulated by Reid~\cite{reid} and the concept of steering as proposed by Wiseman {\it et al.}~\cite{wiseman}. Among the three distinct classes of nonlocal correlations, steering is not symmetric with respect to the observers, unlike entanglement~\cite{ent} and Bell-nonlocality~\cite{nl}, and is also strictly intermediate between 
the latter two~\cite{wiseman}.   Quantification of EPR-steering for two qubit states further brings out the hierarchy between entangled, steerable, Bell-nonlocal states~\cite{costa}.

For the $2-2-2$  (two parties, two measurement settings per party, and two outcomes per measurement)
experimental scenario, the  Bell-Clauser-Horne-Shimony-Holt (Bell-CHSH) inequalities~\cite{CHSH} are the necessary and sufficient conditions for local realism~\cite{fine}. It is well known that all bipartite pure entangled states violate local realist inequalities~\cite{gisin}. This is, however, not true for mixed states~\cite{werner}. A condition on the state parameters for violating the Bell-CHSH inequality by an arbitrary two qubit  state for projective measurements was derived~\cite{horodecki'1995}. 

Though the Reid criterion suffices for demonstrating steering, there
exist several steerable states which do not violate it.  This has motivated the formulation of other or
more optimal steering conditions for both discrete and continuous variables~\cite{entropic1, entropic2, fg1, fg2, gursteer, sumsteer}. It is still an open  problem to find necessary and sufficient conditions for steerability, in general. 
For the $2-2-2$ experimental scenario,  an analogue of the CHSH inequality for steering proposed
recently~\cite{achsh} by Cavalcanti-Foster-Fuwa-Wiseman (CFFW), provides a necessary and sufficient condition for steering. It is thus natural to ask as to how to determine the states that violate
the CFFW inequality.
 
Motivated by the above question in the present work we derive a condition that a two-qubit state
must satisfy in order to violate
the CFFW inequality~\cite{achsh}, {\it a la} the Horodecki condition~\cite{horodecki'1995} for
violation of the Bell-CHSH inequality. Derivation of the criterion for violating the CFFW
inequality enables us to obtain a tight monogamy relation for steering. It is known
that all the three classes of nonlocal correlations are monogamous in nature. Coffman, Kundu and Wootters established quantitatively how entanglement could be shared between more than two 
parties~\cite{ckw}. Later, monogamy relations for quantum violation of the Bell-CHSH inequality for arbitrary tripartite states have been established~\cite{tv}.  Recently, certain monogamy relations for steering have also been proposed~\cite{mreid, vm}.  Steering monogamy relations find applications in threshold efficiency bounds~\cite{efbnd} and one-sided device-independent quantum communication~\cite{banc}.  

The plan of the paper is as follows. In the next section we provide a brief overview of steering
and the CFFW inequality~\cite{achsh}. In section III we derive a criterion for for violating the CFFW inequality 
by an arbitrary bipartite qubit state. The maximal possible violation of this inequality in quantum theory, {\it a la} the Tsirelson bound for Bell nonlocality~\cite{ts}, is then obtained as a corollary of our derivation. As a consequence of this derivation it follows that the CFFW inequality can detect two way steering only, which is not apparent from the expression of the inequality. In section IV
we establish a monogamy relation for the CFFW inequality, which implies that for a tripartite state shared between three observers two of them cannot steer the third party simultaneously through the quantum violation of the CFFW inequality. We end with some concluding remarks in section V.

\section{II. Bipartite steering under two dichotomic measurements per site}

 Steering phenomena can be described as the non-existence of a local hidden variable-local hidden state (LHV-LHS) model for the correlation obtained from local measurements performed on spatially separated systems. Let $A \in \mathbb{M}_{\alpha} $ and $B \in \mathbb{M}_{\beta}$ be the possible choices of measurements for two spatially separated observers, say Alice and Bob, with outcomes $a \in \mathbb{D}_{a}$ and $b \in \mathbb{D}_{b}$, respectively. Alice and Bob perform local measurements independently on a particle in their possession of the bipartite quantum system $\rho_{AB}$. The joint probability of obtaining the outcomes $a$ and $b$, when measurements $A$ and $B$ are performed by Alice and Bob, respectively, is given by, $P(A,B|a,b,\rho_{AB})$.
 
The bipartite state $\rho_{AB}$ of the system is steerable by Alice to Bob iff it is not the case that for all  $A \in \mathbb{M}_{\alpha} $, $B \in \mathbb{M}_{\beta}$, $a \in \mathbb{D}_{a}$, $b \in \mathbb{D}_{b}$, the joint probability distribution can be written in the form
\begin{equation}
P(a, b|A, B, \rho_{AB}) = \sum_{\lambda} P(\lambda )P(a|A,,\lambda)P_Q(b|B,,\rho_{\lambda})
\end{equation}
where $P(\lambda)$ is the probability distribution over the hidden variables $\lambda$,  $P(a|A,\lambda)$ denotes an arbitrary probability distribution, and $P_Q(b|B,\rho_{\lambda})(=tr [\rho_{\lambda}\mathbb{Q}^b_B])$ denotes the quantum probability of outcome $b$ given measurement 
$B$ on the state $\rho_{\lambda}$.

The necessary and sufficient criterion \cite{achsh} to detect steering from Bob to Alice in 
the $2-2-2$ scenario with Alice's measurements being mutually unbiased, is given by
\begin{eqnarray}\label{si}
S_{BA}=\sqrt{\langle (B+B^{\prime})A\rangle^2+\langle (B+B^{\prime})A^{\prime}\rangle^2}+\nonumber\\
\sqrt{\langle (B-B^{\prime})A\rangle^2+\langle (B-B^{\prime})A^{\prime}\rangle^2}\leq 2.
\end{eqnarray}
This inequality, denoted as the CFFW inequality,  is an analog of the Bell-CHSH inequality~\cite{CHSH} which provides a necessary and sufficient condition for two-qubit Bell nonlocality.

\section{III. Condition on two-qubit states for quantum violation of the CFFW inequality}

The Horodecki criterion~\cite{horodecki'1995} decides whether an arbitrary bipartite qubit state violates the CHSH inequality for projective measurements. In a similar spirit here we derive a 
criterion  for the quantum mechanical violation of the CFFW inequality by an arbitrary bipartite qubit state for projective measurements.

An arbitrary state in $\mathcal{H}(=\mathbb{C}^2\otimes\mathbb{C}^2)$ can be expressed in terms of the Hilbert-Schmidt basis as
\begin{eqnarray}\label{rho}
\rho=\frac{1}{4}(\mathbb{I}\otimes\mathbb{I}+\vec{r}.\vec{\sigma}\otimes\mathbb{I}+\mathbb{I}\otimes\vec{s}.\vec{\sigma}+\sum_{i,j=1}^{3}t_{ij}\sigma_i\otimes\sigma_j).
\end{eqnarray}
Here $\mathbb{I}$ is the identity operator acting on $\mathbb{C}^2$, $\sigma_i$s are the three Pauli matrices, and $\vec{r}, \vec{s}$ are vectors in $\mathbb{R}^3$ (with norm less than or equal to unity), with $\vec{r}.\vec{\sigma} = \sum_{i=1}^{3} r_i \sigma_i$, and $\vec{s}.\vec{\sigma} = \sum_{i=1}^{3} s_i \sigma_i$. In addition, for being a valid density matrix, $\rho$ has to be normalised and positive semi-definite. Let us define a matrix $V=TT^t$, where $T$ is the correlation matrix of the state (\ref{rho}) with coefficients $t_{ij} = Tr( \rho \sigma_i \otimes \sigma_j )$. If $v$ and $\tilde{v}$ are the two greatest eigenvalues of $V$, we can define a quantity
\begin{eqnarray}
S(\rho)=\sqrt{v+\tilde{v}}.
\label{srho}
\end{eqnarray}

\emph{Theorem 1}: \textit{There exist projective measurements for which any two-qubit density matrix (\ref{rho}) violates the CFFW inequality (\ref{si}) if  $S(\rho)>1$.}

\emph{Proof}: In order to prove the above theorem let us denote $Q=\hat{q}.\vec{\sigma}$, where $Q\in\{A, A^{\prime}, B, B^{\prime}\}$ and $q\in\{a, a^{\prime}, b, b^{\prime}\}$. The left hand side of the CFFW inequality (\ref{si}) for Bob to Alice steering can be expressed as
\begin{eqnarray}\label{sf}
S_{BA}=\sqrt{\langle (\hat{b}+\hat{b}^{\prime}).\vec{\sigma}\otimes\hat{a}.\vec{\sigma}\rangle^2+\langle (\hat{b}+\hat{b}^{\prime}).\vec{\sigma}\otimes\hat{a}^{\prime}.\vec{\sigma}\rangle^2}+\nonumber\\
\sqrt{\langle (\hat{b}-\hat{b}^{\prime}).\vec{\sigma}\otimes\hat{a}.\vec{\sigma}\rangle^2+\langle (\hat{b}-\hat{b}^{\prime}).\vec{\sigma}\otimes\hat{a}^{\prime}.\vec{\sigma}\rangle^2}.
\end{eqnarray}
We now maximize (\ref{sf}) over all possible projective measurements. To this end, let us define two mutually orthogonal unit vectors $\hat{c}, \hat{c}^{\prime}$ through $\hat{b}+\hat{b^{\prime}}=2\cos\theta\hat{c}$ and $\hat{b}-\hat{b}^{\prime}=2\sin\theta\hat{c}^{\prime}$, with $\theta\in\{0,\pi/2\}$. With the
above definitions, one may write
\begin{eqnarray}\label{msf} 
\underset{\theta, \hat{a}, \hat{a}^{\prime}, \hat{c}, \hat{c}^{\prime}} {Max} S_{BA}= \underset{\theta, \hat{a}, \hat{a}^{\prime}, \hat{c}, \hat{c}^{\prime}} {Max}[2\cos\theta\sqrt{(\hat{c},T\hat{a})^2+(\hat{c},T\hat{a}^{\prime})^2}+\nonumber\\
 2\sin\theta\sqrt{(\hat{c}^{\prime},T\hat{a})^2+(\hat{c}^{\prime},T\hat{a}^{\prime})^2}]\nonumber\\
= \underset{\theta, \hat{a}, \hat{a}^{\prime}, \hat{c}, \hat{c}^{\prime}} {Max} [2\cos\theta\sqrt{(\hat{a},T^t\hat{c})^2+(\hat{a}^{\prime},T^t\hat{c})^2}+\nonumber\\
 2\sin\theta\sqrt{(\hat{a},T^t\hat{c}^{\prime})^2+(\hat{a}^{\prime},T^t\hat{c}^{\prime})^2}].
\end{eqnarray}
Here $(\hat{c},T\hat{a})$ denotes the inner product between $\hat{c}$ and $T\hat{a}$. 

For the CFFW inequality in case of Bob to Alice steering, $\hat{a}$ and $\hat{a}^{\prime}$ are mutually unbiased measurements. Since we restrict ourselves to local projective measurements, $\hat{a}$ and $\hat{a}^{\prime}$  have to be mutually orthogonal~\cite{achsh}. Now, maximizing over $\hat{a}$, and $\hat{a}^{\prime},$ we have
\begin{eqnarray} 
\underset{\hat{c},\hat{c}^{\prime}}{Max} S_{BA} = \underset{\hat{c},\hat{c}^{\prime},\theta}{Max}~~ (2\cos\theta ||T^t\hat{c}||+2\sin\theta ||T^t\hat{c}^{\prime}||)\nonumber\\
=\underset{\hat{c},\hat{c}^{\prime}}{Max}~2\sqrt{||T^t\hat{c}||^2+||T^t\hat{c}^{\prime}||^2} .
\end{eqnarray}
The above expression is maximized when $\hat{c}, \hat{c}^{\prime}$ are the two eigenvectors of $T^t$ corresponding to the two greatest eigenvalues $v, \tilde{v}$. Thus, 
\begin{equation}
\underset{\hat{c},\hat{c}^{\prime}}{Max} S_{BA} = 2\sqrt{v+\tilde{v}}
\label{statecond}
\end{equation}
This value coincides with the Horodecki $M$ function~\cite{horodecki'1995} for the CHSH inequality. Therefore, the quantum mechanical violation of the CFFW inequality for Bob to Alice steering 
occurs when $S_{BA} > 2$, or $S(\rho)>1$.

\emph{Corollary 1:}  There exists local unitary operations such that $T$ can be made diagonal without increasing the nonlocal character of the state. For symmetric matrices, $\lambda_{max}(AB)\leq\lambda_{max}(A)\lambda_{max}(B)$. As $||T||\leq 1$, it follows that $v, \tilde{v}$ can be at most equal to $1$. Hence, the 
maximum value of the left hand side of the CFFW inequality attainable in quantum mechanics is $2\sqrt{2}$.

\emph{Corollary 2:} Note that the maximum value of $S(\rho)$ (\ref{srho}) is independent of the 
steering direction. The expression for $S(\rho)$ remains the
same  by either maximizing the function $S_{BA}$ for Bob to Alice steering, or the corresponding function $S_{AB}$ for Alice to Bob steering. It hence follows that the CFFW inequality detects
two-way steering, a fact that is not evident through the form of the inequality (\ref{si}).\\

It is to be noted that our results are consistent with \cite{parth} where the authors generalised CFFW inequality and shown that states which are EPR steerable with CHSH-type correlations are also Bell nonlocal.

\section{IV. Monogamy of steering}

The restriction on the sharing of quantum correlations of a multipartite system between several number of observers is quantitatively expressed through monogamy relations for entanglement \cite{ckw}, CHSH inequality \cite{tv} and EPR steering \cite{mreid, vm}. Unlike entanglement and Bell-nonlocality, EPR steering is asymmetric with respect to the observers, requiring extra sophistication in the monogamy relations. A monogamy relation through the Reid criterion may be used to quantify the amount of bipartite EPR steering that can be shared by a number of parties, and identify the directionality of steering monogamy~\cite{mreid}. A monogamy constraint  derived using volume of the quantum steering ellipsoid  states that one party cannot steer both of the other two parties to a large set of 
states~\cite{vm}.
 It may be noted that among various monogamy relations, more fundamental are those which are based on some necessary and sufficient criteria. The Reid criterion is not necessary for EPR steering  by
 various continuous variable systems~\cite{entropic2, fg2, gursteer}. Here we derive a monogamy relation for the necessary and sufficient  steering condition in the case of two binary measurements performed by each party~\cite{achsh}.
 
 
\emph{Theorem 2}: For a tripartite state shared between Alice, Bob and Charlie, each having choice to measure between two dichotomic observables, the left hand side of the CFFW inequality for Bob to Alice steering ($S_{BA}$) and that for Charlie to Alice steering ($S_{CA}$), satisfy the relation $S_{BA}^2+S_{CA}^2\leq 8$.

{\it Proof}: To prove this one has to show 
\begin{eqnarray}\label{mono}
\underset{A_i, B_j, C_k}{Max}S_{BA}^2+S_{CA}^2\leq 8,
\end{eqnarray} 
where, $A_i, B_j, C_k$ denote measurements performed by Alice, Bob and Charlie, respectively. The
left hand side of (\ref{mono}) can be written as
\begin{eqnarray}
\underset{A_i, B_j, C_k}{Max}S_{BA}^2+S_{CA}^2\leq \underset{A_i, B_j}{Max}~S_{BA}^2 + \underset{A_i, C_k}{Max}~S_{CA}^2.
\end{eqnarray}
since measurements $A_i$ achieving the maximum in $S_{BA}$ and $S_{CA}$ may differ.
The maximum value of the CHSH expression for a pure three qubit state $\rho$  is given by~\cite{tv}
\begin{eqnarray}
\underset{\hat{c},\hat{c}^{\prime}}{Max}~ 2\sqrt{||T\hat{c}||^2+||T\hat{c}^{\prime}||^2}
=2\sqrt{tr[T_{AB}T_{AB}^t]}
\end{eqnarray}
where $T_{AB}$, the correlation matrix of reduced state belonging to Alice and Bob  is of the form 
\begin{eqnarray}
T_{AB}=\left(\begin{array}{cc}
\langle \sigma_x^A\sigma_x^B\rangle & \langle \sigma_x^A\sigma_z^B\rangle \\ 
\langle \sigma_z^A\sigma_x^B\rangle & \langle \sigma_z^A\sigma_z^B\rangle \\ 
\end{array}\right),
\end{eqnarray}
with $\langle \sigma_x^A\sigma_x^B\rangle =tr[\sigma_x^A\otimes\sigma_x^B\otimes\mathbb{I}\rho]$. Similar results hold for the reduced state of Alice and Charlie.
From theorem 1, it is known that the maximum value of $S_{AB}$  is the same as that of CHSH expression. Therefore, the monogamy relation derived  for the CHSH inequality~\cite{tv} is also applicable for 
the CFFW inequality with the same choice of measurements for Alice in $S_{BA}$ and in $S_{CA}$. 
Hence, one obtains
\begin{eqnarray}\label{smi}
S_{BA}^2+S_{CA}^2 \leq 8.
\end{eqnarray}
The inequality \ref{smi} implies if $S_{BA}>2$ i.e., Bob can steer Alice's subsystem then it can not be steered by an another party. 
This inequality can be saturated for the three qubit state given in Ref.~\cite{tv}, which also saturates the CHSH  monogamy inequality.

\section{V. Conclusions}

Steering is a phenomenon where one party by her choice of measurements can prepare different assemblages of another system not possessed by her, which cannot be mimicked by any classical model. In the 
$2-2-2$ (two parties, two measurement settings per party, and two outcomes per measurement setting) experimental scenario, the necessary and sufficient condition for steering is given by
the  CFFW inequality~\cite{achsh}. Here we have derived a condition on an arbitrary quantum state of two qubits for  the violation of the CFFW inequality  using projective measurements. Our analysis closely follows
the derivation of a similar state condition for the CHSH inequality~\cite{horodecki'1995}. 
The maximum possible quantum violation of the CFFW inequality and its inability to detect one-way steering follow as corollaries of our derived criterion which represents a necessary and sufficient 
condition on the steerability of a quantum state under the above restrictions.
Monogamy of quantum correlations is an important non-classical property which is relevant for tasks 
such as secure key generation between separated parties.  Here we establish a tight monogamy relation for two-qubit steering with two measurements per party.

{\it Acknowledgements}:
DD acknowledges financial support from the University Grants Commission, Govt. of India.

\end{document}